# Enhanced Hydrogen Evolution Activity of MOS2-rGO Composite Synthesized via Hydrothermal Technique.

Abhishek Sebastian [1][0000-0002-3421-1450] and Pragna R [1][0000-0003-0827-5896]

[1] Abhira, Department of Nano Technology, Chennai

**Abstract.** Hydrogen evolution reaction (HER) has emerged as a promising technique for the production of clean and sustainable energy. In recent years, researchers have been exploring various materials for efficient HER activity. In this study, we report the synthesis of two different materials, namely MOS2 and MOS2-rGO, through a hydrothermal technique. X-ray diffraction (XRD), Fourier-transform infrared (FTIR) spectroscopy, and Raman spectroscopy were used to characterize the materials. XRD analysis revealed the formation of hexagonal MOS2 with a high degree of crystallinity. FTIR analysis confirmed the presence of Mo-S bonds, while Raman spectroscopy provided evidence for the formation of MOS2.To evaluate the HER activity of the materials, linear sweep voltammetry (LSV) was performed. The results showed that MOS2 and MOS2-rGO had good HER activity with low onset potentials and high current densities. The MOS2 -rGO material showed improved HER activity compared to MOS2, indicating the potential of graphene oxide as a co-catalyst to enhance the performance of MOS2.

## 1 Introduction

Water electrolysis is a sustainable approach for hydrogen production that involves using electricity from clean energy sources to split water into its constituent elements, hydrogen and oxygen. It is an important process for the production of hydrogen because hydrogen is a promising candidate for use as a renewable and scalable energy carrier.

However, the hydrogen evolution reaction (HER) associated with water electrolysis is kinetically sluggish, which leads to low efficiency in corresponding electrolysis devices. HER is the heart of various energy storage and conservation systems, and has attracted even more attention during the past decades due to the growing demand for renewable energy and the possibility of utilizing fuel cells as green devices for energy conservation.

The required characteristics of HER catalysts include the ability to reduce the overpotential drastically, to take into account the processes of electron transfer, proton diffusion, and bubble release in the design and engineering of the electrode, and to be made of abundant and inexpensive elements. In recent years, numerous transition-metal based catalysts with abundant reserves in the globe, such as chalcogenides, nitrides, phosphides, borides, and carbides, have been extensively explored for HER. Molybdenum disulfide (MOS2) has gained extensive attention among transition-metal dichalcogenide materials due to its excellent physical, chemical, and mechanical properties.

Current promising HER catalysts include Mo and W sulfide materials, which are made of abundant and inexpensive elements and have good catalytic performances. Molybdenum disulfide (MOS2) composite cathodes have electrochemical performance superior to stainless steel or Pt-based cathodes in phosphate or perchlorate electrolytes, yet they cost significantly less than Pt-based composite cathodes. Two promising organic compounds that can catalyze HER are hydrogenase and nitrogenase enzymes. Transitional metal sulfides such as nanometer-scaled MOS2 and WS2 have attracted considerable attention recently as inorganic electrocatalysts for HER.

## 2 Literature Survey

Wen-Hui Hu et al. proposed Ultrathin MOS2-coated carbon nanospheres as highly efficient electrocatalysts for hydrogen evolution reactions [1] where the Ultrathin MOS2-coated-acid-treated carbon nanospheres (MOS2/ATCNS) were prepared by hydrolysis of Ammonium Thiomolybdate in N, N-dimethylformamide. The Conventional three-electrode system was used with Ag/AgCl as the reference electrode and Platinum Foil as the counter electrode to perform Hydrogen Evolution Reaction (HER) wherein according to Linear Sweep Voltammetry results, MOS2/ATCNS proves to be an efficient catalyst for HER

Bo Gao et al. proposed 3D flower-like defected MOS2 magnetron-sputtered on candle soot for enhanced hydrogen evolution reaction [2] where the primary material used is Candle Soot which is hybridized into a 3D flower-like MOS2 on candle soot Substrate by using the technique of Magnetron Sputtering. The Unique Structure and existence of Metallic MOS2 in its 1T phase improve the electrochemical HER activity at a significant rate due to the larger surface area and rich defects which exposes more active edge sites and shortened diffusion channels for charge transfer and separation. The proposed material shows excellent long-term stability while performing electrochemical reactions

Jarin Joyner et al. proposed Graphene Supported MOS2 Structures with High Defect Density for Efficient HER Electrocatalysts [3]. Graphene Oxide flakes were prepared using Hummer Method, reduced to reduced-Graphene Oxide (rGO) composite foams which are in turn annealed under argon. To perform HER, graphite served as the counter electrode, Ag/AgCl as the reference electrode and glassy carbon as the rotating working electrode in the three-electrode set-up. The lowest onset potential was found to be at 197 mV.

Sriram Kumar et al. proposed Electrochemical and SECM Investigation of MOS2/GO and MOS2/rGO Nanocomposite Materials for HER Electrocatalysis [4]. Ammonium Molybdate acts as the raw material which is processed into MOS2/GO or MOS2/rGO (in its 1T phase) mixed in mortar in the ratio of 1:4. High desorptive charge-transfer resistance was observed in the case of MOS2/rGO which resulted in subsequent low HER. On the other hand, the catalytic activity of the composite material for the HER process was observed to be comparatively equal to the commercially used Pt/C catalyst.

Yingying Xiao et al. proposed Ethylenediamine-assisted phase engineering of 1T/2H - MOS2/graphene for efficient and stable electrocatalytic hydrogen evolution [5]. Modified Hummer's method was used to synthesize Graphene Oxide (GO). Hydrothermal method and magnetic stirring were performed to form an evenly dispersed mixture of Mo:S with a molar ratio of 1:8. MOS2/rGO was synthesized to its 1T phase which activates more active sites and improves the electroconductivity. The incorporation of conductive rGO exhibits remarkable HER performance and excellent durability in an acidic environment.

Harish Reddy Inta et al. proposed Ionic Liquid-intercalated Metallic MOS2 as a Superior Electrode for Energy Storage Applications [6]. Sodium Molybdate and thioacetamide were synthesized using the solvothermal method to produce 1T-MOS2@CC and 2H-MOS2@CC. Due to high interlayer spacing and improved conductivity along the basal plane, 1T-MOS2@CC shows higher electrochemical activity towards HER along with reasonable supercapacitor behaviour. It is also observed that direct growth of MOS2 samples on CC results in an increased electrochemical activity compared to that of powder counterparts that were drop cast on CC using binders.

Hoa Thi Bui et al. proposed In-situ formation and integration of graphene into MOS2 interlayer spacing: expansion of interlayer spacing for superior hydrogen evolution reaction in acidic and alkaline electrolytes [7]. Sodium Molybdenum oxide dihydrate and thiourea were used to synthesize MOS2@Gr heterostructures. X-Ray diffraction and High-Resolution Transmission Electron Microscopy techniques revealed that integrated graphene expanded the width of MOS2 interlayers exposing a higher number of active edge sites for enhanced HER. Long-term electrochemical endurance in acidic and alkaline electrolytes against HER was also observed.

Zhexue Chen et al. proposed Size-dependent and support-enhanced electrocatalysis of 2H-MOS2 for hydrogen evolution [8]. Full-Scale Nano sheets of 2H-MOS2 in vari-ous lateral thicknesses were used as the catalyst. Performing Electrochemical Impedance Spectroscopy to observe the reaction kinetics, Cyclic Voltammetry and High-Resolution Transmission Electron Microscopy revealed that MOS2-$Ti_3C_2$ QS/NS heterostructure demonstrates the best electrocatalytic activity against HER.

Wencui Zhang et al. proposed a Superior Hydrogen Evolution Reaction Performance in 2H-MOS2 to that of the 1T Phase [9]. Silicon Wafers, Single Crystal Bulk MOS2 were used as raw materials to synthesize MOS2 in its 2H phase. Images produced by Scanning Electron Microscopy, Energy dispersive X-Ray spectra and Raman Mapping prove that Excellent electrocatalytic activity is observed when 2H-MOS2 is applied with the vertical electric field and modulated under the fermi level.

Kan Zhang et al. proposed Aligned Heterointerface-Induced 1T-MOS2 Monolayer with Near-Ideal Gibbs Free for Stable Hydrogen Evolution Reaction [10]. Dimethyl Sulfoxide (DMSO) was processed using a disproportionate reaction in the presence of (OH-) ions to form dimethyl sulfide which further reduces GO and Mo5+ ions from MoCl5 to form MOS2. 1mL of Sodium Hydroxide was used to increase the point of zero charges to favour a more negative surface charge that helps in anchoring Mo ions and Glassy Carbon. A precursor-in-solvent strategy is proposed to realize 1T-MOS2 stabilization at edge aligned interface of 2H-MOS2 and rGO. It is also observed that the novel strategy producing EA-2H/1T/rGO structure exhibits better HER activity than the usage of regular 1T-MOS2.

Federico Fioravanti et al. proposed the Effect of MOS2 in doped-reduced graphene oxide composites - Enhanced electrocatalysis for HER [11]. Glassy Carbon, rGO flakes, nitrogen atoms and sulfur atoms from caffeine and thiocyanate, aqueous dispersion of MOS2 and Nafion solution are processed together to form a final hybrid of MOS2/N-rGO/GC and MOS2/SN-rGO/GC. The hybrids show great enhancement for HER due to the expansion degree of S-Mo-S bonds. These MOS2-based hybrids are also useful for electrochemical sensors. It is concluded that HER modulations are based on the nature of the heteroatom and the graphenic structure along with its interaction towards MOS2. The best performance towards HER was observed in acidic media obtained from MOS2/SN-rGO with the precise amount of interstitial S atoms between 2 graphene layers as it increases the rate of electron transfer.

Parisa Salarizadeh et al. proposed MOS2 coating on different carbonaceous materials: Comparison of electrochemical properties and hydrogen evolution reaction performance [12]. Molybdenum di sulfide, Carbonaceous materials such as Reduced Graphene Oxide (rGO), Hollow Carbon Nanosphere (HCN), Single-walled Carbon Nanotubes (SWCNTs), Multi-walled Carbon Nano Tubes (MWCNT), Sodium Molybdate, Thioacetamide and Glassy Carbon were processed to produce Hybrid materials like MOS2/rGO, MOS2/ HCNS, MOS2/SWCNT, MOS2/MWCNT. The catalysts coated on Glassy Carbon act as the working electrode, Ag/AgCl acts as the reference electrode and Pt wire act as the working electrode. Due to the higher conductivity of rGO and the efficient electronic coupling between rGO and MOS2, the volume of hydrogen that evolved for the hybrid MOS2/rGO is high when compared to the other hybrids. Out of the 4 hybrids, MOS2/rGO is observed to have the highest electrocatalytic efficiency and also requires the least onset overpotential with a value of 25mV and a comparatively low tafel slope with a value of 44mV/dev.

Tran Chien Dang et al. proposed MOS2 hydrogen evolution catalysis on p-Si nanorod photocathodes [13]. p-Si (100) wafers (Nanorods) were coated with MOS2 flakes using Metal-organic Chemical Vapor Deposition Method (MOCVD), Acetone, Methanol, Hydroflurioic acid, Silver Nitrate and Hydrogen Peroxide were used to clean the substrate and Au/Pt/Ti wafers were used for annealing. The fabricated materials also have supercapacitor applications. The resultant layers of MOS2 flakes coated on P-Si Nanorods are a highly potential catalyst for PEC applications promoting the approaches to produce heterojunction materials to enhance Hydrogen Evolution Efficiency.

Yuelong Xu et al. proposed the Facile fabrication of molybdenum compounds (Mo2C, MoP and MOS2) nanoclusters supported on N-doped reduced graphene oxide for highly efficient hydrogen evolution reaction over a broad pH range[14]. Ammonium Tetra-thio-molybdate, 1,3,5-Trisbenzoic acid, and 2-amino-benzoic acid were processed to produce P-rGO/Mo as raw material which is further synthesized to produce rGO/MOS2. Polyvinylpyrrolidone is used as a radical indicator, and Ammonia and Hydrochloric acid are used as shielding gases. Electrocatalytic performance of Mo2C, MoP and MOS2 were compared with the prepared samples investigated in 1.0M KOH and 0.5M H2SO4 solution. N-doped porous rGO with MOS2 which was prepared using peroxide-assisted step and hydrothermal step method was observed to have plentiful defects and porous structure that provided active sites to facilitate the catalysis. It is concluded that the work provides a strategy to prepare porous materials for electrocatalysts in the future.

The following papers have provided great insights on the current trends of HER with MOS2 and its hybrids.

# 3 METHODOLOGY

## 3.1 Preparation of MOS2

The synthesis of MOS2 is a process that involves the use of a hydrothermal method. This method is commonly used for the synthesis of various inorganic materials due to its ability to produce high-quality crystalline structures. The process of synthesizing MOS2 using this method involves the use of two precursors, sodium molybdate dihydrate and thiourea, which were purchased from a reputable supplier.

The first step in the synthesis of MOS2 involves the dispersion of the molybdenum precursor in distilled water. The molybdenum precursor was added to the water in a specific ratio, which was determined through a series of experiments. The ratio of mo-lybdenum precursor to thiourea used in this study was 1:4. The mixture was then agitated for 15 minutes to ensure homogeneity. Homogeneity is a crucial factor in the success of the synthesis process, as it ensures that the precursors are evenly distributed throughout the mixture.

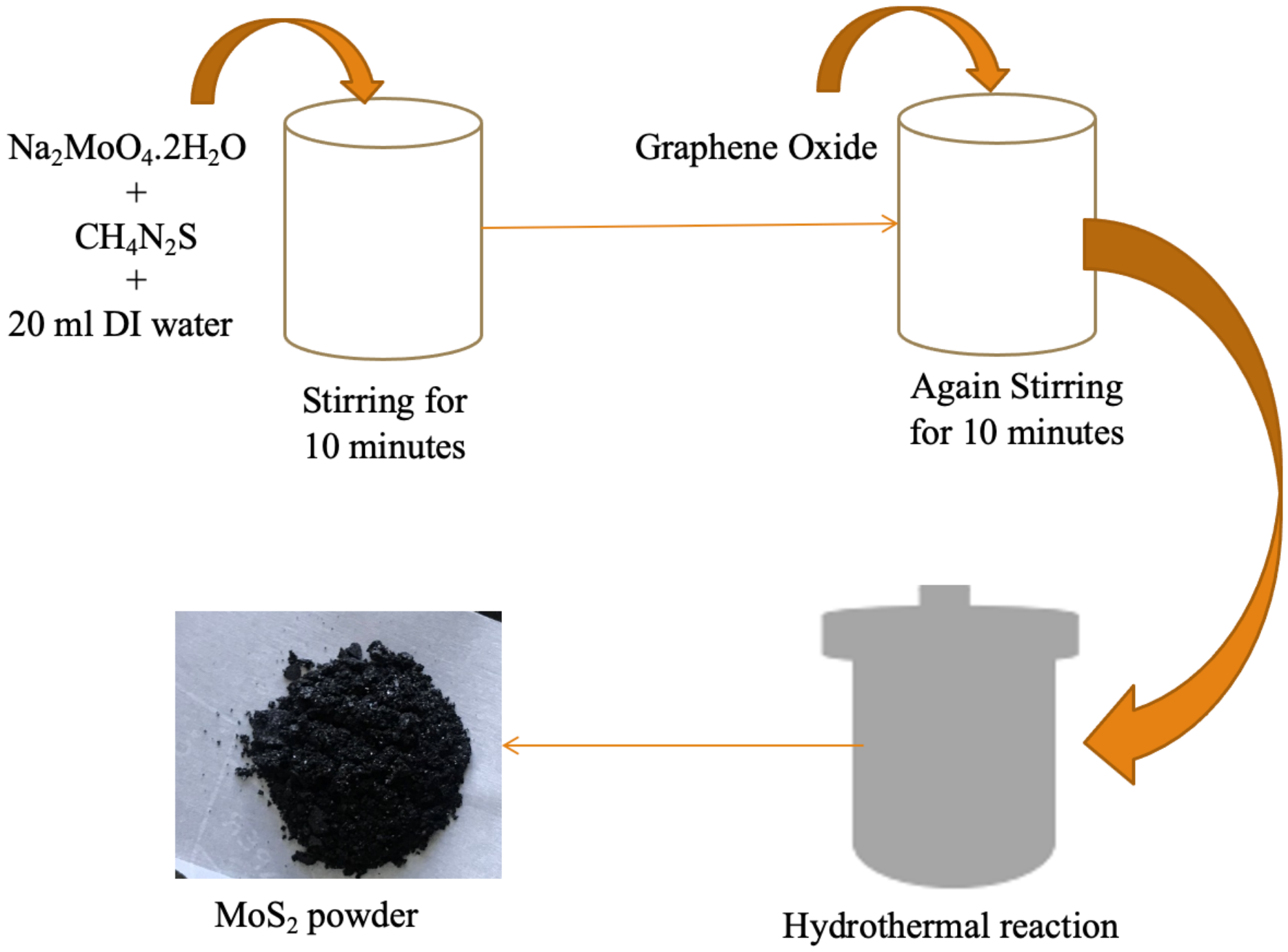

Figure1. Synthesis of MOS2 Using Hydrothermal Route

Once the mixture was homogenized, it was placed in a Teflon-coated stainless-steel autoclave. The autoclave was designed to withstand high temperatures and pressures, making it ideal for the hydrothermal method. The autoclave was heated to a temperature of 200°C and maintained at this temperature for 22 hours. The temperature and time of the heating process were determined based on the results of previous experiments and were found to be optimal for the synthesis of MOS2.

After the heating process was completed, the autoclave was cooled to room temperature. The cooling process is an essential step as it allows the synthesized material to solidify and form crystals. Once the autoclave had reached room temperature, the powders were removed and dried for 8 hours at a temperature of 80°C in an oven. The drying process is necessary as it removes any residual water or other impurities from the synthesized material, ensuring that the final product is of high quality.

The synthesis of MOS2 using the hydrothermal method is a complex process that requires careful attention to detail. The success of the process depends on several factors, including the homogeneity of the mixture, the temperature and time of the heating process, and the cooling and drying processes. These factors were carefully controlled in this study to ensure that the synthesized MOS2 was of high quality and suitable for use in various applications.

# 4 EXPERIMENTS AND RESULTS

## 4.1 Characterizations of MOS2 and MOS2-rGO:

X-ray diffraction (XRD), Fourier-transform infrared (FTIR) spectroscopy, and Raman spectroscopy are common techniques used for the characterization of materials. These techniques provide important information about the crystal structure, chemical composition, and bonding of materials. In the case of MOS2 and MOS2-rGO, XRD can be used to analyze the crystalline structure of the materials, while FTIR and Raman spectroscopy can provide insights into their chemical composition and bonding.

XRD is a non-destructive analytical technique that is widely used for the identification of crystal structures. In XRD, a sample is exposed to X-rays, and the diffraction pattern produced is analyzed to determine the crystal structure of the material. For MOS2 and MOS2-rGO, XRD can be used to determine the crystal structure and phase of the materials, as well as to identify any impurities or defects in the crystal structure.

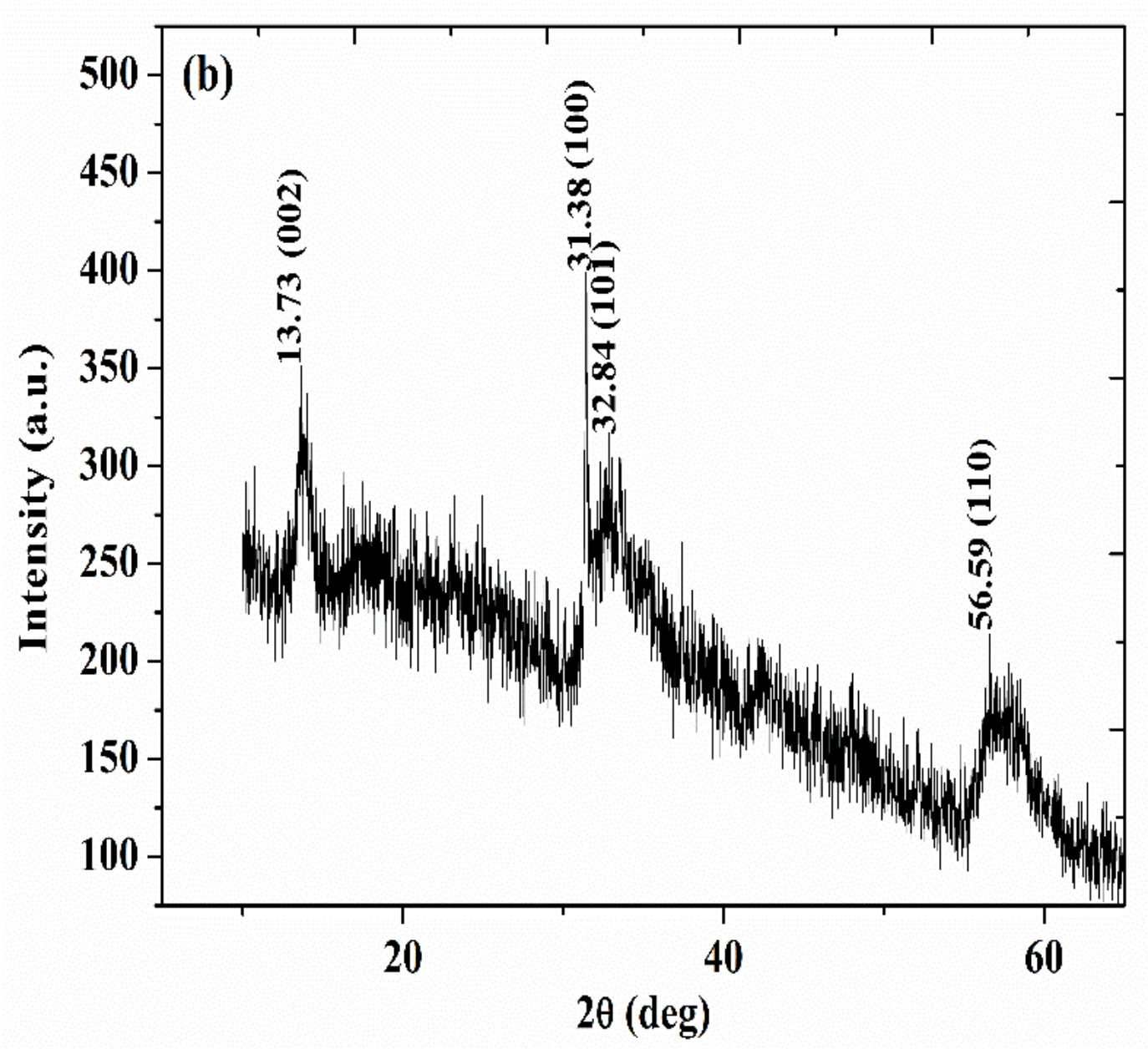

Figure 2: XRD Characterization of MOS2

From Figure 2, We can Infer that Well-defined broad XRD peaks at 2θ values of 13.73°, 31.38°, 32.84° and 56.59° are obtained corresponding to (002), (100), (101) and (110) planes of 2H-MOS2, which is well matched with standard JCPDS card number 37-1492

FTIR spectroscopy, on the other hand, is a powerful technique for analyzing the chemical bonding and functional groups in a material. In FTIR, a sample is exposed to infrared radiation, and the absorption and transmission of the radiation are measured. This provides information about the vibrational modes of the molecules in the sample, which can be used to identify the functional groups and chemical bonding in the material. FTIR can be used to analyze the functional groups and bonding in MOS2 and MOS2-rGO, which can provide insights into their chemical properties and reactivity.

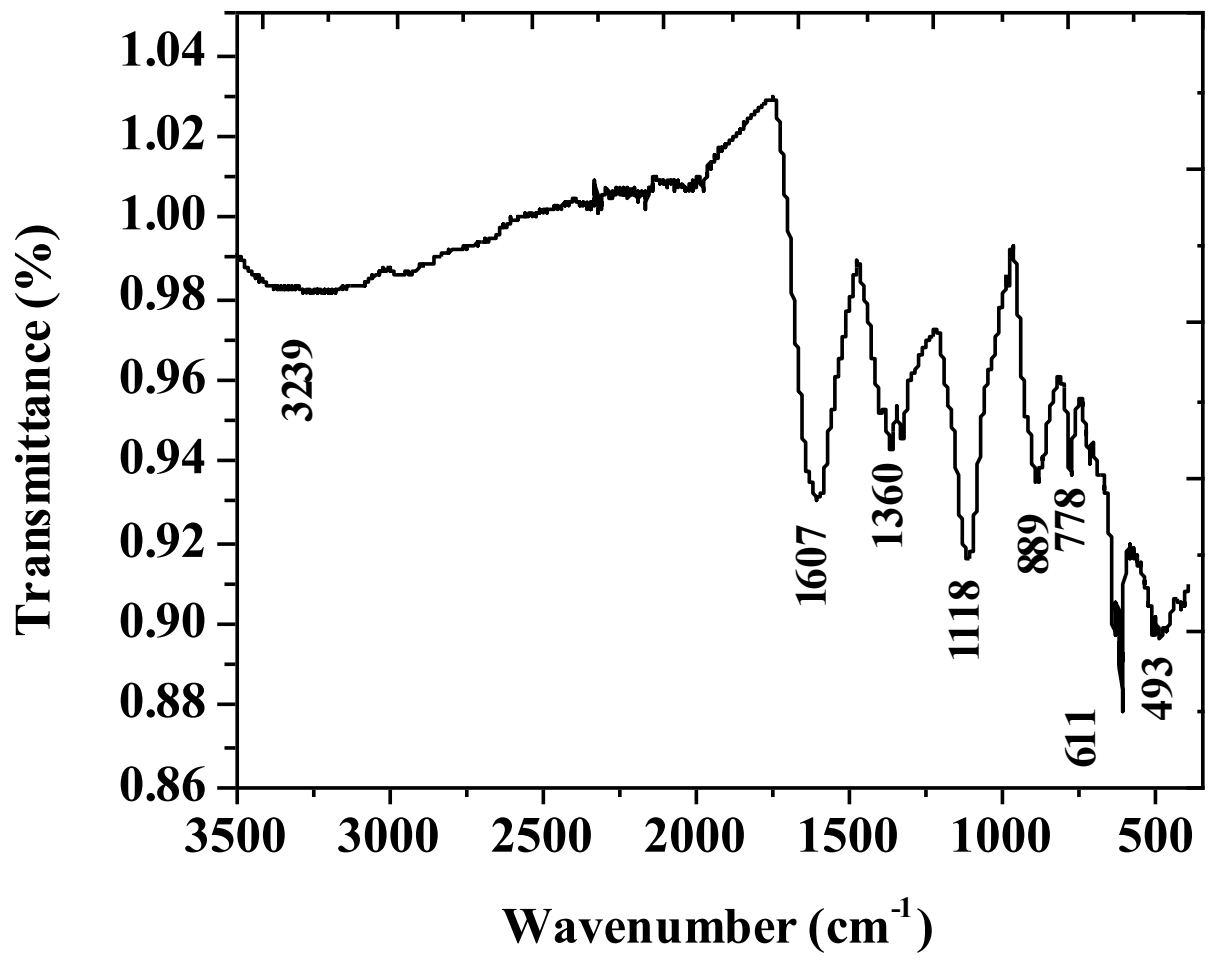

Figure 3: FTIR Characterization of MOS2

From Figure 3, We can infer that the Absorption bands at 889 cm−1, 1360 cm−1, and 1607 cm−1, which are attributed to MOS2, the band at 493 cm−1 is due to the S–S bond, Absorbance peaks of S=O and Mo-O were observed at wavelengths 1118 and 611 cm-1, respectively, The peaks at about 3239 cm−1 belong to the characteristic bands of the O–H group.

Raman spectroscopy is another technique used for the characterization of materials. Raman spectroscopy measures the scattering of monochromatic light by a sample, which can be used to determine the vibrational modes of the molecules in the sample. This provides information about the chemical bonding and structure of the material. Raman spectroscopy can be used to analyze the bonding and structure of MOS2 and MOS2-rGO, and can provide insights into their electronic and optical properties.

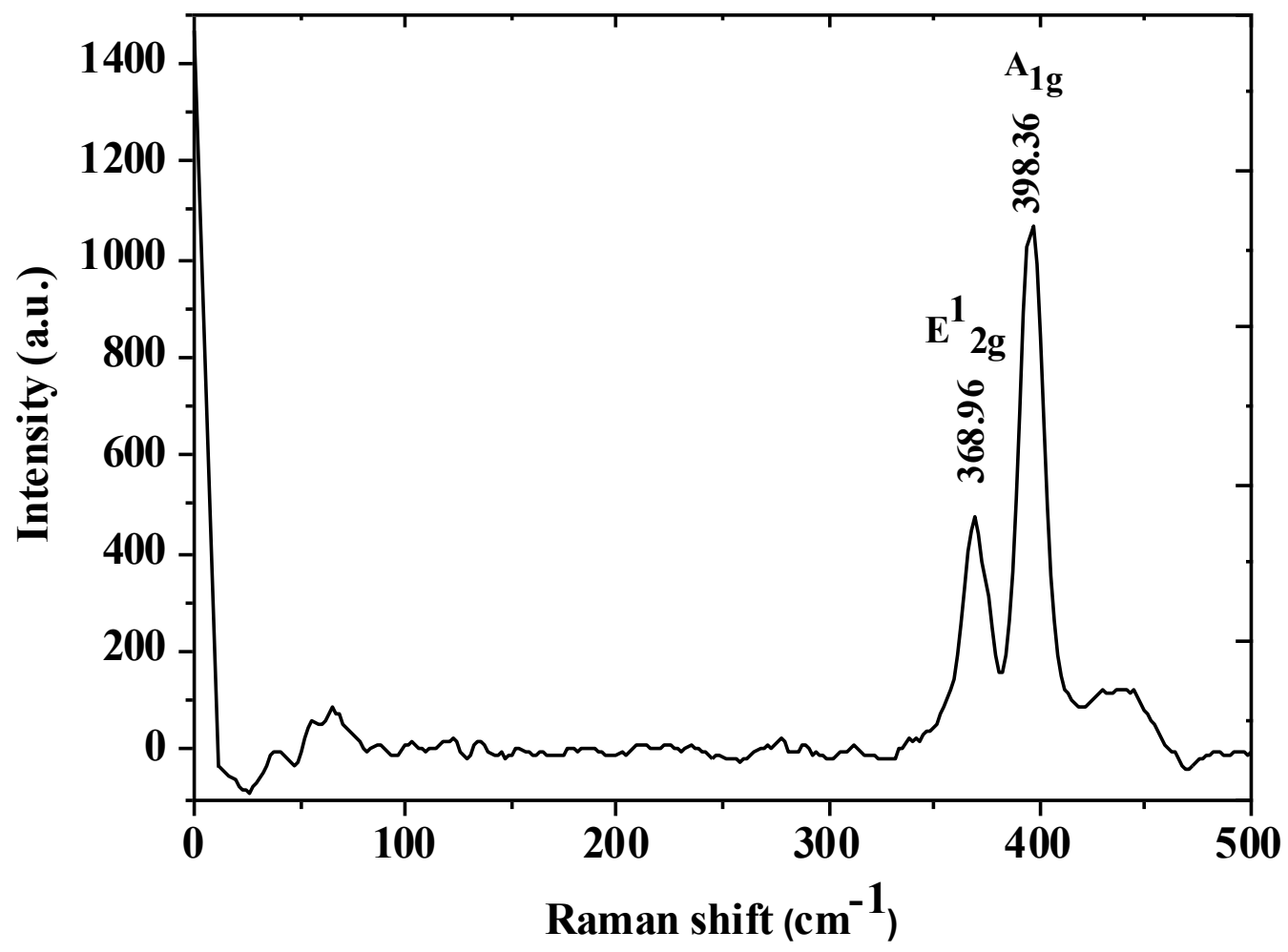

Figure 4: Raman Spectroscopy Characterization of MOS2

From Figure 4, Raman spectrum shows the dominant $E^12g$ in-plane and $A_{1g}$ out-of-plane modes are observed at 368.96 cm−1 and 398.36 cm−1 corresponding to pure 2H-MOS2 phase.

It Can be inferred that; with the results of the following characterizations the synthesized material is MOS2

## 4.2 Linear Sweep Voltammetry

Linear Sweep Voltammetry (LSV) is a commonly used electrochemical technique that can be used to study the kinetics of various electrochemical reactions, including the hydrogen evolution reaction (HER). The HER is an important reaction in electrocatalysis, as it is a crucial step in the production of hydrogen, a clean and renewable energy source. In recent years, there has been increasing interest in the use of two-dimensional materials such as MOS2 and MOS2-rGO as electrocatalysts for the HER. LSV can be used to evaluate the electrocatalytic activity of these materials for the HER by measuring the current as a function of the applied potential. By studying the LSV curves, it is possible to extract important information about the kinetics of the HER on these materials, such as the onset potential, the Tafel slope, and the exchange current density.

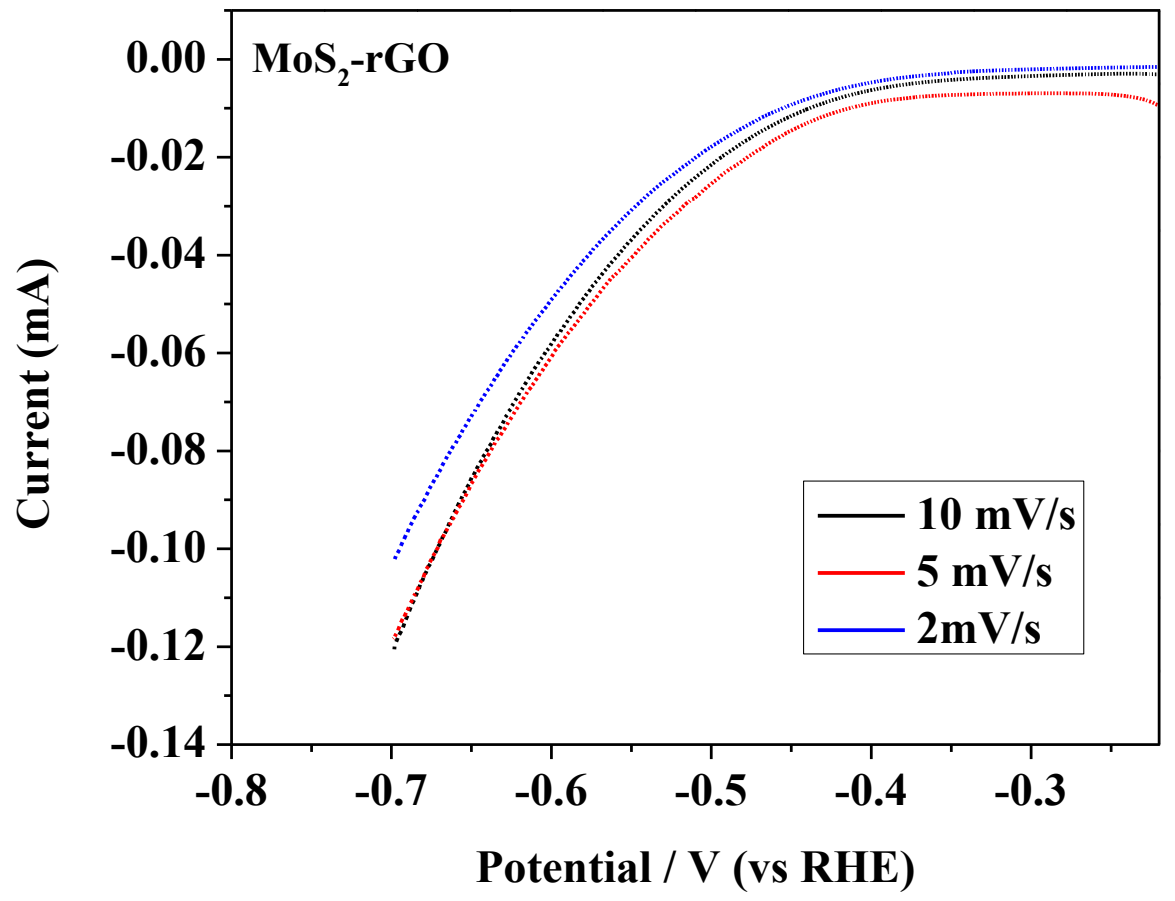

Figure 5. Linear sweep voltammetry curves of pristine 2H-MOS2 for different scan rates

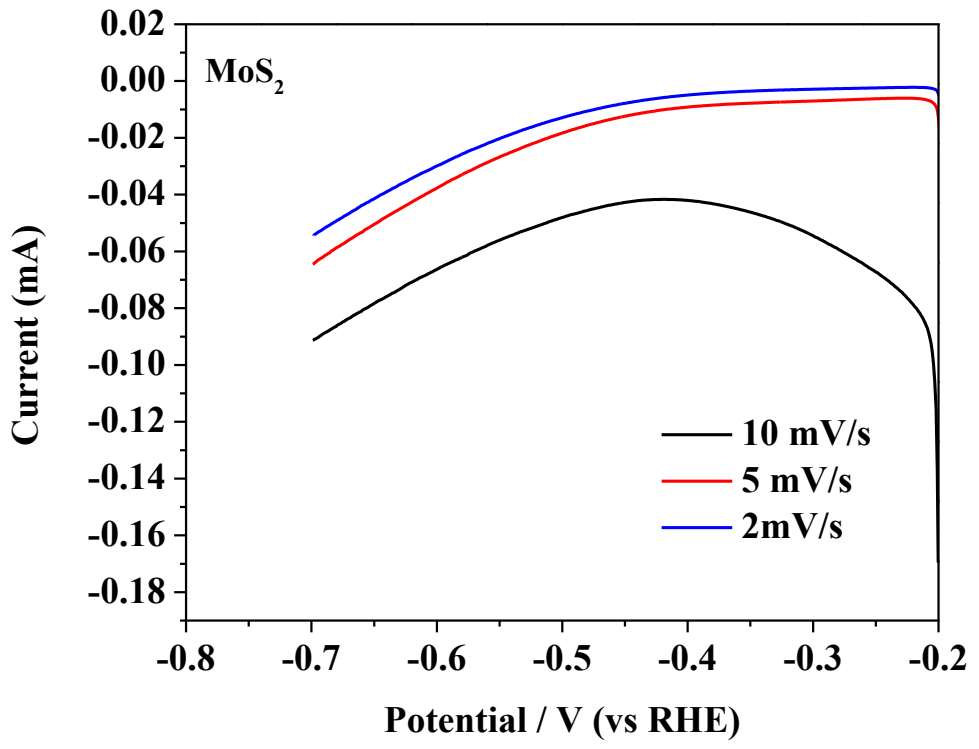

Figure 6. Linear sweep voltammetry curves of MOS2/rGO composite for different scan rates

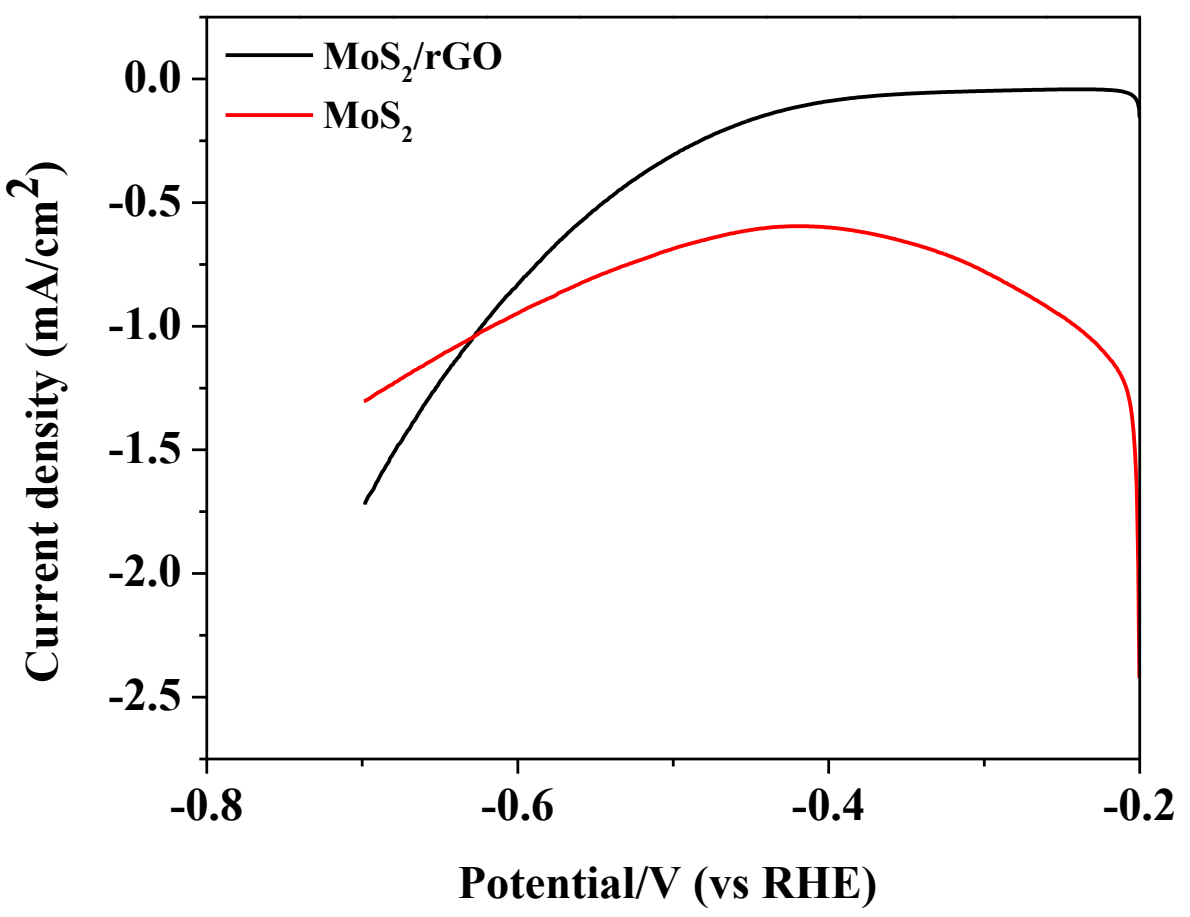

Figure 7. Linear sweep voltammetry curves for 2H-MOS2 and MOS2/rGO for 10 mV/s scan rate.

The LSV graphs obtained for MOS2 and MOS2 -rGO show that the addition of GO to MOS2 leads to a significant improvement in the electrocatalytic activity towards HER. The HER performance is characterized by the onset potential, which is the potential at which the current density starts to increase significantly, and the overpotential, which is the potential required to reach a certain current density. The LSV graphs for MOS2 -rGO show a lower onset potential and overpotential compared to MOS2, indicating improved electrocatalytic activity. This suggests that the addition of GO enhances the electron transfer between MOS2 and the electrolyte, leading to improved HER performance. Furthermore, the LSV graphs also show a larger cathodic current density for MOS2 -rGO, indicating an increased rate of hydrogen evolution. Overall, the LSV results strongly support the idea that the incorporation of GO into MOS2 enhances its electrocatalytic performance towards HER.

# 5 CONCLUSION

In conclusion, water electrolysis is a sustainable approach for hydrogen production and has gained significant attention in recent years due to the growing demand for renewable energy. The hydrogen evolution reaction (HER) associated with water electrolysis is kinetically sluggish, leading to low efficiency in corresponding electrolysis devices. To improve the efficiency of HER, numerous transition-metal based catalysts, such as molybdenum disulfide (MOS2) have been extensively explored for their catalytic properties.

In this study, a hydrothermal method was used to synthesize MOS2, which was characterized using XRD, FTIR, and Raman spectroscopy techniques. The synthesized MOS2 was of high quality and suitable for use in various applications. The use of MOS2-rGO composites has also been explored as a promising HER catalyst due to their excellent electrochemical performance and cost-effectiveness.

Overall, this study provides important insights into the synthesis and characterization of MOS2 and MOS2-rGO composites, which have promising applications in the field of renewable energy and HER catalysts. Further research is needed to explore the full potential of these materials and to improve their performance in HER applications.